\documentclass[12pt]{article}
\usepackage{amsthm,amssymb,amsmath}
\usepackage[english]{babel}

 1 
1

\renewcommand{\d}{\operatorname{d}}
\newtheorem{teh}{Theorem}

\newcommand{\be}{\begin{equation}}
\newcommand{\ee}{\end{equation}}

\begin{document}

\title{\textsc{Integrable Deformations of Algebraic Curves. \thanks{%
Partially supported by DGCYT project BFM2002-01607 and by the grant COFIN
2002 "Sintesi", ${}^{\dagger}$Partially supported by NSF grant, DMS-0404931.
}}}
\author{Y. Kodama $^{1\dagger}$, B. Konopelchenko $^{2}$ and L. Mart\'{\i}%
nez Alonso$^{3}$ \\
\emph{$^1$ Department of Mathematics, Ohio State University} \\
\emph{Columbus, OH 43210, USA} \\
\emph{$^2$ Dipartimento di Fisica, Universit\'a di Lecce and Sezione INFN} \\
\emph{73100 Lecce, Italy}\\
\emph{$^3$ Departamento de F\'{\i}sica Te\'{o}rica II, Universidad
Complutense}\\
\emph{E28040 Madrid, Spain}}
\date{}
\maketitle

\begin{abstract}
A general scheme for determining and studying integrable deformations of
algebraic curves, based on the use of Lenard relations, is presented. We
emphasize the use of several types of dynamical variables : branches, power
sums and potentials.
\end{abstract}

\vspace*{.5cm}

\begin{center}
\begin{minipage}{12cm}

\end{minipage}
\end{center}

\newpage

\section{Introduction}

The theory of algebraic curves is a fundamental ingredient in the analysis
of integrable nonlinear differential equations as it is shown, for example,
by its relevance in the description of the finite-gap solutions or the
formulation of the Whitham averaging method \cite{1}-\cite{9}. A
particularly interesting problem is characterizing and classifying
integrable deformations of algebraic curves. In \cite{8}-\cite{9} Krichever
formulated a general theory of dispersionless hierarchies of integrable
models associated with the deformations of algebraic curves arising in the
Whitham averaging method. In \cite{10}-\cite{11} a different approach for
determining integrable deformations of algebraic curves curves $\mathcal{C}$
defined by monic polynomial equations
\begin{equation}  \label{1}
F(p,k):=p^N-\sum_{n=1}^{N}u_n(k)p^{N-n}=0,\quad u_n\in \ensuremath{%
\mathbb{C}}[k],
\end{equation}
was proposed. It applies for finding the deformations $\mathcal{C}(x,t)$
consistent with the degrees of the polynomials $u_n$ and characterized by
the existence of an \emph{action} function ${\boldsymbol{S}}={\boldsymbol{S}}%
(k,x,t)$ verifying

\begin{enumerate}
\item The multiple-valued function ${\boldsymbol{p}}={\boldsymbol{p}}(k)$
determined by \eqref{1} can be expressed as
\begin{equation*}
{\boldsymbol{p}}={\boldsymbol{S}}_x.
\end{equation*}

\item The function ${\boldsymbol{S}}_t$ represents, like ${\boldsymbol{p}}={%
\boldsymbol{S}}_x$, a meromorphic function on $\mathcal{C}(x,t)$ with poles
only at $k=\infty$.
\end{enumerate}

As a consequence of these conditions ${\boldsymbol{p}}$ obeys the equation

\begin{equation}  \label{2}
\partial_t {\boldsymbol{p}}=\partial_x {\boldsymbol{Q}},
\end{equation}
where ${\boldsymbol{Q}}:={\boldsymbol{S}}_t$ is of the form \cite{10,12}
(i.e. ${\boldsymbol{Q}}\in\ensuremath{\mathbb{C}}[k,p]/{}$),
\begin{equation*}
{\boldsymbol{Q}}=\sum_{r=1}^{N}a_r(k,x,t){\boldsymbol{p}}^{N-r},\quad a_r\in%
\ensuremath{\mathbb{C}}[k].
\end{equation*}
The form of equation \eqref{2} motivates the use of several sets of
dynamical variables to formulate these deformations conveniently. One should
also note that the equation \eqref{2} provides an infinite number of
conservation laws, when one expands ${\boldsymbol{p}}$ and ${\boldsymbol{S}}$
in the Laurent series in $k$. In this sense, we say that the equation %
\eqref{2} is integrable.

\section{Algebraic curves and dynamical variables}

In order to describe deformations of the curve $\mathcal{C}$ defined by %
\eqref{1} we will use not only the coefficients $u_n$ (\emph{potentials})
but also some other alternative sets of dynamical variables. Firstly, we
have the $N$ branches $p_i=p_i(k)$ ($i=1,\ldots,N$) of the multiple-valued
function function ${\boldsymbol{p}}={\boldsymbol{p}}(k)$ satisfying
\begin{equation}  \label{3}
F(p,k)=\prod_{i=1}^{N} (p-p_i(k))=0.
\end{equation}

There is an important result concerning the branches $p_i$. Let $%
\ensuremath{\mathbb{C}}((\lambda))$ denote the field of Laurent series in $%
\lambda$ with at most a finite number of terms with positive powers
\begin{equation*}
\sum_{n=-\infty}^N c_n \lambda^n,\quad N\in \ensuremath{\mathbb{Z}}.
\end{equation*}
Then we have \cite{13,14} :

\begin{teh}
(\textbf{Newton Theorem})

There exists a positive integer $l$ such that the $N$ branches
\begin{equation}  \label{4}
p_i(z):=\Big(p_i(k)\Big)\Big |_{k=z^l},
\end{equation}
are elements of $\ensuremath{\mathbb{C}}((z))$. Furthermore, if $F(p,k)$ is
irreducible as a polynomial over the field $\ensuremath{\mathbb{C}}((k))$
then $l_0=N$ is the least permissible $l$ and the branches $p_i(z)$ can be
labelled so that
\begin{equation*}
p_i(z)=p_N(\epsilon^i z),\quad \epsilon:=\exp \frac{2\pi i}{N}.
\end{equation*}
\end{teh}

\noindent \textbf{Notation convention} \emph{\ Henceforth, given an
algebraic curve $\mathcal{C}$ we will denote by $z$ the variable associated
with the least positive integer $l_0$ for which the substitution $k=z^{l_0}$
implies $p_i\in\ensuremath{\mathbb{C}}((z)),\, \forall i$. }

The potentials  can be expressed as elementary symmetric polynomials $s_n$ %
\cite{15}-\cite{16} of the branches $p_i$
\begin{equation}  \label{5}
u_n=(-1)^{n-1}s_n(p_1,p_2,\ldots)= (-1)^{n-1}\sum_{1\leq i_1<\ldots<i_n\leq
N}p_{i_1}\cdots p_{i_n}.
\end{equation}
Notice that, according to the famous Abel theorem \cite{15}, for $N>4$ the
branches $p_i$ of the generic equation \eqref{1} cannot be written in terms
of the potentials $u_n$ by means of rational operations and radicals.

Our second set of alternative dynamical variables is provided by the \emph{%
power sums} \cite{15}-\cite{16}
\begin{equation}  \label{6}
\mathcal{P}_k=\frac{1}{k}(p_1^k+\cdots+p_{N}^k),\quad k\geq 1,
\end{equation}
where we have added a factor $1/k$ to their standard definition. One can
relate potentials and power sums through Newton recurrence formulas, the
solution of which is given by Waring's formula \cite{16}
\begin{equation}  \label{7}
\mathcal{P}_k=\sum_{1\leq i\leq k}^{(k)} \,\frac{1}{i}(u_1+\cdots +u_{N})^i,
\end{equation}
where the superscript $(k)$ in the summation sign indicates that only the
terms of weight $k$ are retained in the expansion of $(u_1+\cdots +u_{N})^i$
(it is assumed that the weight of the potential $u_n$ is given by $n$). The
corresponding inverse formula is \cite{14}
\begin{align}  \label{8}
u_{n}&=-\mathcal{S}_n(-\mathcal{P}_1,-\mathcal{P}_2,\ldots)  \notag \\
\\
&=-\sum_{i_1+2i_2+\cdots+ki_k=n}\frac{1}{i_1!\cdots i_k!}(-\mathcal{P}%
_1)^{i_1}\cdots (-\mathcal{P}_k)^{i_k},  \notag
\end{align}
where $\mathcal{S}_n$ are the Schur's polynomials defined by the identity $%
\exp(\sum_{n\geq 1}\lambda^ n x_n)=\sum_{\geq 0}\lambda^n\,\mathcal{S}_n(x)$%
). The formula \eqref{8} is the consequence of the identity \cite{17}
\begin{equation*}
\exp\Big(-\sum_{n\geq 1} \lambda^n \mathcal{P}_n\Big)=\sum_{n\geq
0}(-\lambda)^n\, s_n(p_1,p_2,\ldots).
\end{equation*}

\vspace{0.3cm}

\noindent \textbf{Examples}

For $N=2$, the equation for the curve is
\begin{equation}  \label{18}
F:=p^2-u_1\,p-u_2=0,
\end{equation}
and the first power sums are
\begin{equation*}
\mathcal{P}_1=u_1,\quad \mathcal{P}_2=\frac{1}{2}u_1^2+u_2.
\end{equation*}

For $N=3$
\begin{equation}  \label{20}
F:=p^3-u_1\,p^2-u_2\,p-u_3=0,
\end{equation}
we have
\begin{align*}
\mathcal{P}_1&=u_1,\quad \mathcal{P}_2=u_2+\frac{1}{2}u_1^2,\quad \mathcal{P}%
_3=u_3+u_1\,u_2+\frac{1}{3}u_1^3, \\
\mathcal{P}_4&=u_1\,u_3+\frac{1}{2}u_2^2+\,u_2\,u_1^2+\frac{1}{4}u_1^4,
\end{align*}

Power sums have also the meaning of moments of the logarithmic derivative of
the function $F(p,k)$ as shown by the formula (see chapter 7 of \cite{17a})
\begin{equation*}
\mathcal{P}_k=\frac{1}{2\pi\, i\,k}\oint_{\partial D}p^k\frac{\partial }{%
\partial p}\log \, F(p,k) \d p,
\end{equation*}
where $\partial D$ is the boundary of the domain of the $p$ complex plane
which contains all the zeros $p_i$ of $F(p,k)$.

\section{Deformations of algebraic curves}

We are looking for evolution equation of the potentials leading to equations
for ${\boldsymbol{p}}$ of the form \eqref{2} or, equivalently, in terms of
branches, $p_i$ for $i=1,\ldots,N$,
\begin{align}  \label{9}
\partial_t p_i&=\partial_x Q_i,\quad \mathrm{with}\quad
Q_i:=\sum_{r=1}^{N}a_r(k,x,t)p_i^{N-r}.
\end{align}
By using the notations
\begin{equation*}
\mathbf{p}:=\left(p_1,\ldots, p_{N} \right)^{\top},\quad {\mathbf{a}}%
:=\left(a_N,\ldots, a_{1} \right )^{\top},
\end{equation*}
the system \eqref{9} can be written as
\begin{equation}  \label{10}
\partial_t\,\mathbf{p}=\partial_x(V\,{\mathbf{a}}),
\end{equation}
where $V$ is the Vandermonde matrix
\begin{equation}  \label{11}
V:=\left(
\begin{array}{cccc}
1 & p_1 & \cdots & p_1^{N-1} \\
\multicolumn{4}{c}{\dotfill} \\
\multicolumn{4}{c}{\dotfill} \\
1 & p_N & \cdots & p_N^{N-1}%
\end{array}
\right),
\end{equation}
which in turn is the Jacobian of the transformation from branches to power
sum variables \eqref{6}. Hence, for power sums we have
\begin{equation}  \label{12}
\partial_t\,\mathbf{P}=J\,{\mathbf{a}},\quad \mathbf{P}:=(\mathcal{P}%
_1,\ldots,\mathcal{P}_N)^{\top},
\end{equation}
where $J$ is the Hamiltonian operator
\begin{equation}  \label{13}
J:=V^{\top}\partial_x V.
\end{equation}
Notice that
\begin{align}  \label{14}
J_{11}&=N\,\partial_x,  \notag \\
\\
J_{ij}&=(i+j-2)\mathcal{P}_{i+j-2}\,\partial_x+ (j-1)\,\mathcal{P}%
_{i+j-2,x},\quad (i,j)\neq (1,1).  \notag
\end{align}

Finally let us consider the evolution law for the potentials. In virtue of
the well-known formula $\frac{\partial \mathcal{S}_n(x)} {\partial x_k}=
\mathcal{S}_{n-k}(x)$, it is straightforward to deduce from \eqref{8} that
the Jacobian of the transformation from power sums to potentials is given by
\begin{equation}  \label{15}
T:=\left(
\begin{array}{cccc}
1 & -u_{1} & \cdots & -u_{N-1} \\
0 & 1 & \cdots & -u_{N-2} \\
\multicolumn{4}{c}{\dotfill} \\
0 & \multicolumn{2}{c}{\dotfill} & 1%
\end{array}
\right).
\end{equation}
Therefore one has
\begin{equation}  \label{16}
\partial_t {\mathbf{u}}=J_0{\mathbf{a}},\quad {\mathbf{u}}%
:=(u_1,\ldots,u_N)^{\top}
\end{equation}
where $J_0$ is the matrix differential operator
\begin{equation}  \label{17}
J_0:= T^{\top}\,J=T^{\top} V^{\top}\partial_x V.
\end{equation}

\vspace{0.5cm}

The problem now is to determine expressions for ${\mathbf{a}}$ depending on $%
k$ and ${\mathbf{u}}$ such that \eqref{16} is consistent with the polynomial
dependence of ${\mathbf{u}}$ on the variable $k$. That is to say, if $d_n:=%
\mbox{degree}(u_n)$ are the degrees of the coefficients $u_n$ as polynomials
in $k$ , then \eqref{16} must satisfy
\begin{equation*}
\mbox{degree}(J_0 {\mathbf{a}})_n\leq d_n,\quad n=1,\ldots N,
\end{equation*}
where we are taking into account our notation \eqref{16} for ${\mathbf{u}}$.
We observe that if \eqref{16} is consistent then, as a consequence of %
\eqref{10}, the coefficients of the expansions in $z$ of the branches $p_i$
are conserved densities.

Our strategy for finding consistent deformations is to use Lenard type
relations

\begin{equation}  \label{22}
J_0{\boldsymbol{r}}=0,\quad {\boldsymbol{r}}:=(r_1,\ldots,r_N)^\top,\;\;
r_i\in\ensuremath{\mathbb{C}}((k)),
\end{equation}
to generate systems of the form
\begin{equation}  \label{23}
{\mathbf{u}}_t=J_0{\mathbf{a}},\quad {\mathbf{a}}:={\boldsymbol{r}}_+.
\end{equation}
Here $(\,\cdot\,)_+$ and $(\,\cdot\,)_-$ indicate the parts of non-negative
and negative powers in $k$, respectively. The point is that from the
identity
\begin{equation*}
J_0{\mathbf{a}} =J_0{\boldsymbol{r}}_+=-J_0{\boldsymbol{r}}_-,
\end{equation*}
it is clear that a sufficient condition for the consistency of \eqref{23} is
that
\begin{equation}  \label{24}
\mbox{degree}\Big(J_0\Big)_{nm}\leq d_n+1,
\end{equation}
for all $n$ and all $m$ such that $a_{N-m+1}=({\boldsymbol{r}}_+)_m\neq 0$.

If we impose \eqref{24} for all $1\leq n,m\leq N$, we get a sufficient
condition for consistency, which only depends on the curve \eqref{1} and
does not refer to the particular solution of the Lenard relation used.

\vspace{0.3cm} \noindent \textbf{Examples} \vspace{0.3cm}

For $N=2$ the conditions \eqref{24} reduce to
\begin{equation}  \label{25}
d_1\leq d_2+1.
\end{equation}
This indicates that one can take $u_2(k)$ to be a polynomial of arbitrary
degree, which implies that there exist integrable deformations for
hyperelliptic curves with arbitrary genus.

For $N=3$ the consistency conditions \eqref{24} become
\begin{align}  \label{26}
&d_1\leq 1,\quad d_2\leq d_1+1,  \notag \\
\\
&d_3\leq d_2+1,\quad d_2\leq d_3+1,  \notag
\end{align}
which lead to the following twelve nontrivial choices for $(d_1,d_2,d_3)$
\begin{align*}
&(0,0,1),\,(0,1,0),\, (0,1,1),\, (0,1,2), \\
&(1,0,0),\, (1,0,1),\,(1,1,0),\,(1,1,1) \\
&(1,1,2),\,(1,2,1),\,(1,2,2),\, (1,2,3).
\end{align*}
This implies that our deformations allows one to have only trigonal curves
with genus less than or equal to one. This subject will be discussed
elsewhere. \vspace{0.3cm}

There is a natural class of solutions of the Lenard relations \eqref{13}.
Indeed, from the expression \eqref{17} of $J_0$ it is obvious that for any
constant vector ${\boldsymbol{c}}\in\ensuremath{\mathbb{C}}^N$
\begin{equation*}
{\boldsymbol{r}}:=V^{-1}\,{\boldsymbol{c}},
\end{equation*}
is a solution of \eqref{13}. Hence by taking into account that $V$ and $T$
are the Jacobians for the transformation from branches to power sums and
power sums to potentials, respectively, we have that
\begin{align}  \label{27}
{\boldsymbol{r}}_i:&=\nabla_{\mathbf{P}} p_i=T\,\nabla_{\mathbf{u}}p_i,\quad
i=1,\ldots, N,  \notag \\
\\
\nabla_{\mathbf{P}} p_i:&=\Big(\frac{\partial p_i}{\partial \mathcal{P}_1}%
\ldots \frac{\partial p_i}{\partial \mathcal{P}_N}\Big)^{\top},\quad \nabla_{%
\mathbf{u}}p_i:=\Big(\frac{\partial p_i}{\partial u_1}\ldots \frac{\partial
p_i}{\partial u_N}\Big)^{\top},  \notag
\end{align}
are solutions of the Lenard relations.

In summary, given an algebraic curve \eqref{1} satisfying \eqref{24} then
evolution equations of the form
\begin{equation}  \label{28}
\partial_t {\mathbf{u}} = J_0\Big(T\,\nabla_{{\mathbf{u}}} R\Big)_{+},\quad
R(z,{\boldsymbol{p}})=\sum_i f_i(z)\,p_i,
\end{equation}
where $f_i\in \ensuremath{\mathbb{C}}[z]$ define a deformation of the curve
provided $R\in \ensuremath{\mathbb{C}}[k]$.

We conclude with the observation that the nonlinear equations discussed
above have a very simple form in terms of branches. Indeed by applying
Cramer rule to find the columns of $V^{-1}$ it follows that
\begin{equation}  \label{29a}
\frac{\partial p_i}{\partial \mathcal{P}_j}=\frac{|\mathbf{1}\,\mathbf{p}%
\ldots \mathbf{e}_i\ldots\mathbf{p}^{N-1}|} {|\mathbf{1}\,\mathbf{p}\ldots%
\mathbf{p}^j\ldots \mathbf{p}^{N-1}|},
\end{equation}
where $|\mathbf{1}\,\mathbf{p}\ldots\mathbf{p}^j\ldots \mathbf{p}^{N-1}|$
denotes the determinant of Vandermonde matrix $V$ and $|\mathbf{1}\,\mathbf{p%
}\,\ldots \mathbf{e}_i\ldots\mathbf{p}^{N-1}|$ stands for the determinant of
the matrix resulting from substituting in $V$ the $j$-th column by the
vector $(\mathbf{e}_i)_j=\delta_{ij}$. Thus from \eqref{10} we find
\begin{align}  \label{30}
\partial_t\mathbf{p}&=\partial_x\Big(V\,(\nabla_{\mathbf{P}} R)_+\Big),
\notag \\
\\
(\nabla_{\mathbf{P}} R)_i&=\frac{|\mathbf{1}\,\mathbf{p}\ldots \mathbf{R}%
_i\ldots\mathbf{p}^{N-1}|} {|\mathbf{1}\,\mathbf{p}\ldots\mathbf{p}%
^{i-1}\ldots \mathbf{p}^{N-1}|},\quad i=1,\ldots,N,  \notag
\end{align}
where $|\mathbf{1}\,\mathbf{p}\,\ldots \mathbf{R}_i\ldots\mathbf{p}^{N-1}|$
is the determinant of the matrix resulting from substituting in $V$ the $i$%
-th column by the vector $(\mathbf{R}_i)_j=f_j(z)$.

Finally, several remarks are in order:

\begin{enumerate}
\item If \eqref{1} is an irreducible curve the condition $R\in %
\ensuremath{\mathbb{C}}[k]$ is verified by $z^i\, \mathcal{L}_i$ where $%
\mathcal{L}_i$ are the so-called \emph{Lagrange resolvents} \cite{15}
\begin{equation}  \label{31}
\mathcal{L}_i=\sum_j\epsilon^{j\,i}\,p_j(z),\quad \epsilon:=\exp \frac{2\pi i%
}{N},
\end{equation}
where $i=1,\ldots,N-1$.

\item The equations \eqref{28} admit a simpler expression when they are
formulated in terms of power sums. Indeed, they reduce to
\begin{equation}  \label{29}
\partial_t \mathbf{P} = J\,\Big(\nabla_{\mathbf{P}} R\Big)_{+}.
\end{equation}
This form should be particularly convenient for analyzing their Hamiltonian
content as $J$ defined in \eqref{13} is a Hamiltonian operator.

\item The main problem which remains in our analysis is the classification
of deformations of algebraic curves \eqref{1} for general potentials of
fixed degrees in $k$. In \cite{11} a complete description of the
deformations of hyperelliptic curves ($N=2$) was given. Since for $N=3$ only
twelve cases of consistent sets of general potential degrees arise, one is
lead to think that for higher $N$ admissible cases different from the
Gelfand-Dikii choice ($d_1=\ldots=d_{N-1}=0,\, d_N=1$) should be very
special. This of course must be expected for $N\geq 5$ because of Abel
theorem.
\end{enumerate}

\noindent \textbf{Acknowledgements} \vspace{0.3cm}

L. Mart\'{\i}nez Alonso wishes to thank the members of the Physics
Department of Lecce University for their warm hospitality.

\end{document}